\documentclass[prc,floatfix,superscriptaddress,showpacs,twocolumn,amssymb,amsmath,amsfonts,aps]{revtex4}
\usepackage{graphicx} 
\usepackage{dcolumn}  
\usepackage{longtable}
\usepackage{bm}       
\usepackage{subfigure}
\usepackage{float}
\usepackage{multirow} %

\newcommand{\piN}{\pi N}
\newcommand{\pipN}{\pi^+ n}
\newcommand{\ksig}{K^+ \Sigma^0}
\newcommand{\klam}{K^+ \Lambda}
\newcommand{\peta}{\eta p}
\newcommand{\petap}{\eta' p}
\newcommand{\pphi}{\phi p}
\newcommand{\pomega}{\omega p}
\newcommand{\kcos}{\cos \theta^{K^+}_{\mbox{\scriptsize c.m.}}}

\begin{document}

\author{Biplab Dey} 
\affiliation{Carnegie Mellon University, Pittsburgh, Pennsylvania 15213}
\author{Curtis A. Meyer} 
\affiliation{Carnegie Mellon University, Pittsburgh, Pennsylvania 15213}
\date{\today}

\title{Normalization discrepancies in photoproduction reactions}

%
%
\begin{abstract} 
Recent CLAS photoproduction results using a tagged bremsstrahlung photon beam for the ground-state pseudoscalar meson photoproduction channels ($K^+ \Lambda$, $K^+ \Sigma^0$, $\eta p$, $\pi^+ n$ and $\pi^0 p$) show a normalization discrepancy with older results from SLAC, DESY and CEA that used an untagged bremsstrahlung beam. The CLAS results are roughly a factor of two smaller than the older data. The CLAS $K^+\Lambda$ and $K^+\Sigma^0$ results are in excellent agreement with the latest LEPS results that also employed a tagged beam. For the vector meson ($\omega p$ and $\phi p$) channels, CLAS agrees with SLAC results that employed a linearly polarized beam using laser back-scattering, as well as Daresbury data that also came from tagged photon experiment. We perform a global survey of these normalization issues and stress on their significant effect on the coupling constants used in various partial wave analyses.   
\end{abstract}

\pacs{11.80.Et,25.20.-x,12.40.Nn} 

\maketitle

\section{Introduction}

Unravelling the spectrum of excited baryons resonances is a fundamental goal of hadronic physics~\cite{klempt}. Dedicated facilities at Jefferson Lab, MAMI, LEPS and elsewhere are currently collecting a wealth of new data using novel polarization experiments that will soon enable us to reconstruct the ``complete'' quantum-mechanical amplitude for photo-hadron reaction mechanisms (see Ref.~\cite{mypolpaper} for a discussion). The final goal is to understand the effective degrees of freedom inside nuclear matter, and thereby, the strong interaction in the non-perturbative limit. At the simplest level, one can interpret the nucleon resonances as excitations of a basic three-quark system (SU(6) spin-flavor symmetry) that are bound by a confining harmonic potential (O(3) radial excitations). In this so-called constitent quark model (CQM), the baryon spectrum emerges as the super-multiplets of the full SU(6)$\times$O(3) symmetry. However, this symmetric CQM model predicts many more (roughly four times) states as seen in $\piN$ partial-wave analyses, leading to the question --  where are the ``missing'' baryon states? In the diquark model, first proposed by Lichtenberg~\cite{lichtenberg}, two out of the three quarks remain clustered together, leading to lesser degrees of freedom and therefore, lesser number of excited states. However, work by Koniuk and Isgur~\cite{koniuk_isgur}, and later, Capstick and Roberts~\cite{capstick_roberts}, have shown that many of these ``missing'' have weak couplings to the $\piN$ sector. Instead, the missing states couple strongly to the non-$\piN$ sector. Since most of the world data exist in the pion channels, this could explain why these missing states have not been seen as yet. In view of these Capstick-Roberts predictions, several high quality datasets for the non-$\piN$ ground meson photoproduction channels $\klam$~\cite{prc_klam}, $\ksig$~\cite{prc_ksig}, $\peta/\petap$~\cite{prc_eta}, $\pomega$~\cite{prc_omega} and $\pphi$~\cite{prc_phi} have been recently published by the CLAS Collaboration from the high-statistics ``g11a'' experiment~\cite{clas_runperiods}. The g11a results extend upon results from a previous lower-energy ``g1c'' dataset~\cite{clas_runperiods} and in the regions of kinematic overlap, the two datasets are found to be in excellent agreement with each other. It seems, however, that a persistent normalization discrepancy exists between CLAS and some other world datasets. To wit, the CLAS differential cross sections for the pseudo-scalar meson channels ($\klam$~\cite{prc_klam,bradford-dcs}, $\ksig$~\cite{prc_ksig,bradford-dcs}, $\peta$~\cite{prc_eta}, $\pi^0 p$~\cite{dugger_pi0N}, $\pi^+ n$~\cite{dugger_pipN}) are systematically lower that those from older high-energy and forward meson production-angle SLAC/DESY/CEA data. On the other hand, the CLAS $\klam$/$\ksig$ results agree well with the latest LEPS forward-angle data~\cite{leps_klam_ksig_sumihama}, although CLAS is lower than CB-ELSA~\cite{crede_eta_2009, crede_eta_2005} for the $\peta$ channel. Also, the CLAS vector-meson ($\pomega$~\cite{prc_omega}, $\pphi$~\cite{prc_phi}) cross sections are in good agreement with both SLAC~\cite{ballam} and Daresbury~\cite{barber_omega,barber_phi} data.

Given that the quoted systematic uncertainties in most places are of the order of $\sim 10\%$, a discrepancy as large as a factor of two is a matter of concern and any partial wave analysis (PWA) based on these data is bound to be affected by this. The effect of these discrepancies on PWAs has already been discussed by Sibirtsev {\em et al.}~\cite{sibirtsev_eta_discrepancy} for the $\peta$ sector. In this work we present a systematic global overview of these discrepancies taking into account several different channels from the CLAS results and provide detailed descriptions of the various internal checks that have been performed. Next, using a Regge-based formalism, we show how much the different hadrodynamic coupling constants are affected depending on which dataset is being fit to. For the pseudoscalar channels, in contrast to the recent work by Yu {\em et al.}~\cite{yu_pion,yu_kaon} who have conjectured the relevance of tensor exchanges to ``resolve'' this discrepancy, we show that once the CLAS g11a data is taken into consideration, it becomes clear that {\em it is impossible to reconcile the old SLAC data and the CLAS data within a single fit}. We round up our discussion by commenting on the possible resolution of this discrepancy using ongoing and future analyses and experiments.

\section{The CLAS ``g11a'' and ``g1c'' experiments}

In this section we give a brief description of the two experiments that the CLAS data draw results from. The ``g11a'' experiment~\cite{clas_runperiods} (2004) was a high-statistics ($\sim 20$ billion event triggers were recorded) experiment dedicated to search for the exotic pentaquark state. Due to the rather specific physics aim that required detection of multiple charged particles in the final state, a ``two-prong'' trigger was used here. The CLAS detector is divided into six azimuthally-symmetric ``sectors''. A two-prong trigger required at least two charged-tracks in two different sectors to be detected in coincidence with an incoming photon within a time window of 150~ns. This trigger setting also meant that single pion channels were not accessible in this experiment. The photons were produced via bremsstrahlung from a 4.023-GeV electron beam and were energy tagged by measuring the 3-momenta of the recoiling electron. The resulting tagged photon energy range ($E_\gamma$) was from 0.808 to 3.811~GeV. Detailed description of CLAS and its various sub-components can be found in Ref.~\cite{mecking} and references therein. Converting $E_\gamma$ to the center-of-mass (c.m.) energy $\sqrt{s}$, g11a was able to make measurements for each channel concerned from near-threshold till $\sqrt{s} \approx 2.84$~GeV. As of this writing, for the pseudo-scalar channels, this represents the highest energy world data that utilized a tagged photon beam.

Since g11a was a high-precision experiment designed to search for an exotic particle, the data underwent an extensive calibration by several groups within CLAS working independently during this process. Sophisticated analysis tools such as a dedicated g11a kinematic fitter~\cite{kin_fitter} was developed to add to the robustness of the results. Care was also taken to keep the data analysis cuts to be as loose as possible. An event-based signal-background separation method~\cite{jinst_williams} and a physics-weighted detector acceptance calculation from a unbinned maximum likelihood partial wave analysis fit ensured that all correlations present in the data were faihfully represented during the yield extraction and acceptance calculation procedures. It is also worth noting at this point that all the six g11a results of concern here ($\klam$~\cite{prc_klam}, $\ksig$~\cite{prc_ksig}, $\peta/\petap$~\cite{prc_eta}, $\pomega$~\cite{prc_omega} and $\pphi$~\cite{prc_phi}) utilized the same set of data analysis tools and photon flux nomalizations, which should keep the systematic uncertainties under better control, as well.

\begin{table}
  \centering
  \begin{tabular}{|l|c|c|} \hline \hline
    \multirow{2}{*}{Characterestics} \ & \multicolumn{2}{|c|}{\;\;\;CLAS experiment\;\;\;} \\
    \cline{2-3}
    & \;\; g11a \;\; & \;\;g1c \;\;\\ \hline
    Run year & 2004 & 1999 \\ \hline
    Cryotarget length & 17.85 cm & 40 cm \\ \hline
    Start counter & new  &  older \\ \hline
    Trigger setting & 2-prong & 1-prong \\ \hline
    $\pi N$-channels accessible? \;\;\;& no & yes \\ \hline
    Kinematic fitter used?& yes & no \\ \hline
    Physics-weighted acceptance?& yes & no \\ \hline
    Maximum $\sqrt{s}$ (GeV) & 2.84 & 2.55 \\ \hline
    Tagged photon beam?& yes & yes \\ \hline
    \hline
    \end{tabular}
  \caption[]{\label{table:g11a_g1c} Some characterestics of the two CLAS experiments of concern in this work. The two datasets had several distinguishing features and were analyzed by completely different groups. However, in the regions of kinematic overlap, the results from g11a and g1c were in excellent agreement with each other.
}
\end{table}

The ``g1c'' experiment~\cite{clas_runperiods} was an earlier (1999) photoproduction experiment that also utilized a tagged beam facility but ran at a lower photon energy. The maximum $\sqrt{s}$ accessible here was $\sim 2.55$~GeV. The other distinction from g11a was that g1c utilized a less-restrictive single-prong trigger setting that allowed it to analyze the single pion channels ($\pi^0 p$~\cite{dugger_pi0N} and $\pi^+ n$~\cite{dugger_pipN}) as well. Compared to g11a, g1c had completely different target characterestics, trigger settings and analysis personnel. Several of the sophisticated analysis tools that g11a used (the kinematic fitter, for example) were also not available at the time of g1c. A flat phase-space Monte Carlo generator was used in the case of g1c, in contrast to a physics-weighted acceptance calculation that was used in g11a. Therefore, although the same CLAS detector was employed in both cases, by and large, the two sets of results can be termed as ``independent''. However, {\em in the regions where kinematics overlapped, the two CLAS experiments were in excellent agreement with each other, lending support to the internal consistency of the CLAS results}.

\section{The discrepancies}

\subsection{$\klam$ and $\ksig$}
\label{sec:klam_ksig_discrepancy}

\begin{figure}
  \includegraphics[width=3.3in]{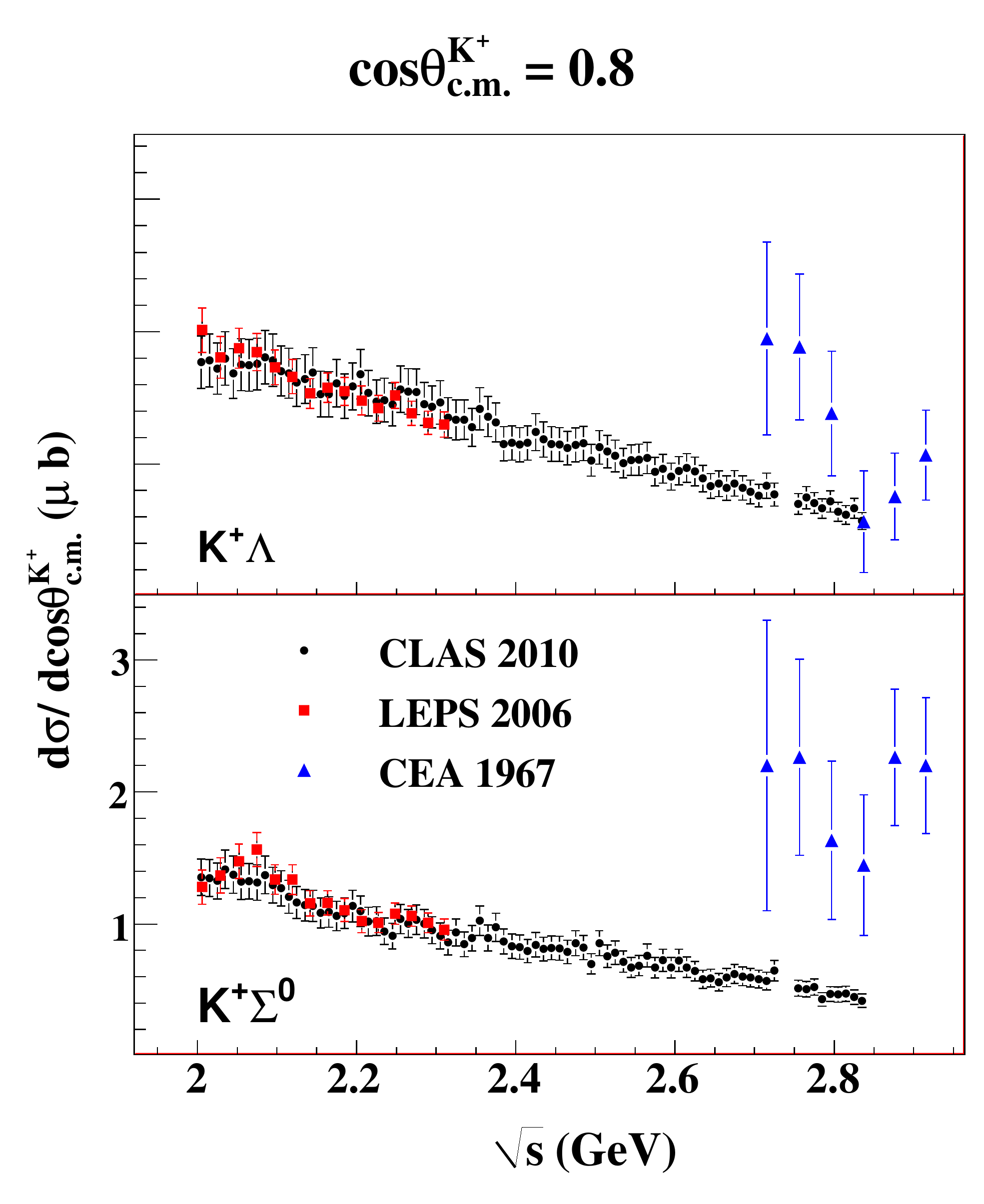}
  \caption[]{\label{fig:klam_ksig_clas_leps_cea} (Color online) Comparison between the CLAS g11a~\cite{prc_klam,prc_ksig}, CEA-Elings-1967~\cite{elings} and LEPS-Sumihama-2006~\cite{leps_klam_ksig_sumihama} results for a forward-angle bin in the hyperon channels. While CLAS and LEPS are in excellent agreement with each other, the older CEA data is systematically higher. The error bars represent the statistical and systematic uncertainties added in quadrature.}
\end{figure}

\begin{figure}
  \includegraphics[width=3.3in]{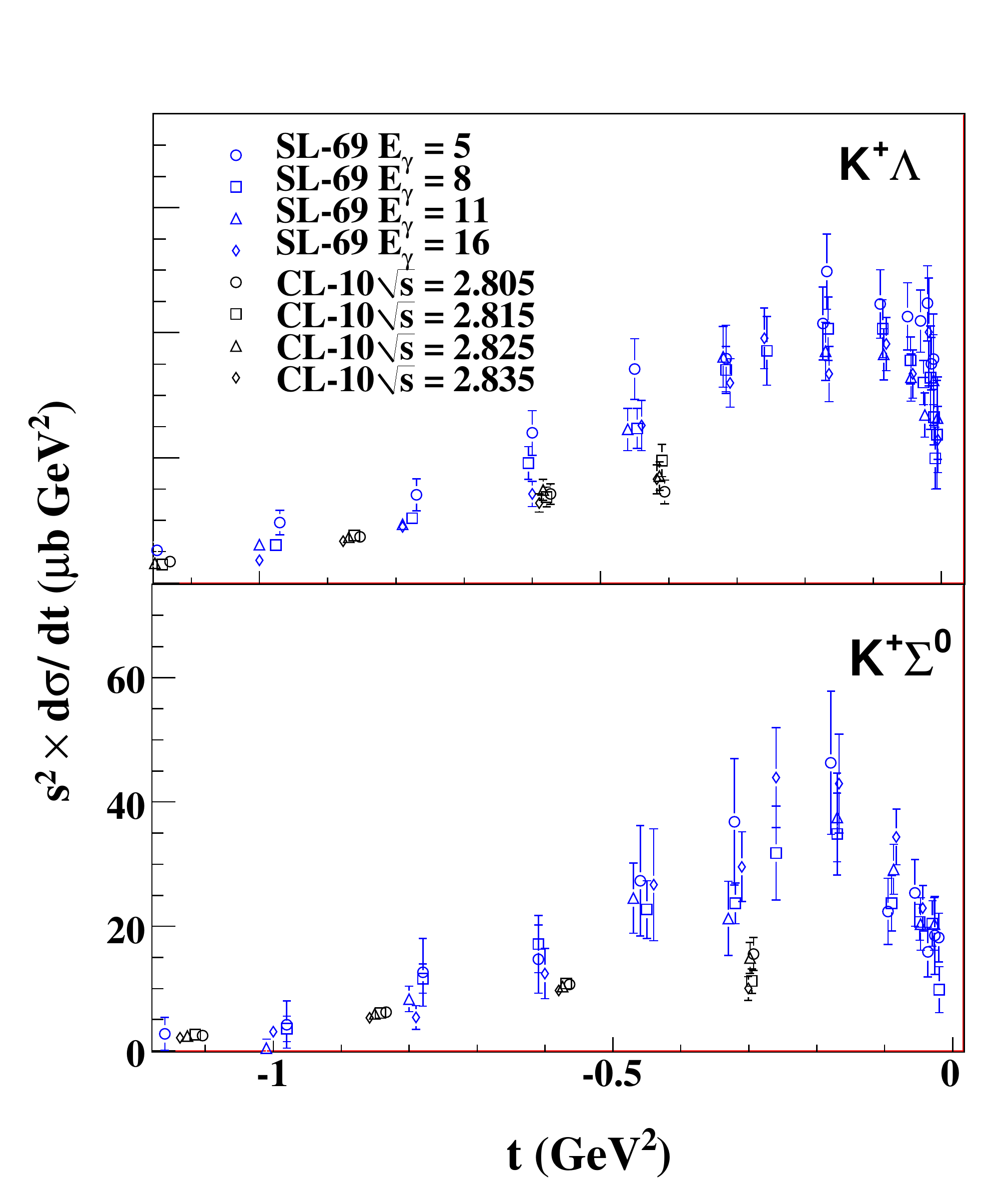}
  \caption[]{\label{fig:clas_slac_scaled_klam_ksig} (Color online) Comparison between the CLAS g11a~\cite{prc_klam,prc_ksig} (CL-10) and SLAC-Boyarski-1969~\cite{boyarski_hyp} (SL-69) scaled cross section results for the two hyperons. The energies are listed in $E_\gamma$ for SLAC and $\sqrt{s}$ for CLAS, in units of GeV. Both sets of results (SLAC and CLAS) agree to a $d\sigma/dt \propto 1/s^2$ behavior as depicted by the tightly bunched set of points, but differ by a normalization factor. The error bars represent the statistical and systematic uncertainties added in quadrature.}
\end{figure}

In what follows, we will broadly refer to the following as the ``SLAC results'' for the two hyperon channels: SLAC-Boyarski-1969~\cite{boyarski_hyp}, CEA-Elings-1967~\cite{elings}, CEA-Joseph-1967~\cite{joseph} and SLAC-Quinn-1979~\cite{quinn}. All of these results covered the high-energy ($E_\gamma > 3$~GeV) forward production angle (small $|t|$) regime and between themselves, agree quite well with each other. These results also exhibit a $t$-channel Regge-type $d\sigma/dt \propto 1/s^2$ scaling behavior. On the other hand, none of these results had a tagged photon beam, but quoted the photon-energy $E_\gamma$ as the end-point energy of the bremsstrahlung spectrum. As was already pointed out in the CLAS g11a $\ksig$ paper~\cite{prc_ksig}, the CLAS results also agree well to a $d\sigma/dt \propto 1/s^2$ Regge-type behavior. Therefore, the CLAS and SLAC data agree well in {\em shape}. However, the CLAS cross sections are systematically lower than SLAC in {\em scale} by roughly a factor of two.

Unfortunately, the kinematics of two sets of results ({\em i.e.} CLAS and SLAC) do not overlap much, which makes a direct comparison somewhat difficult. The SLAC results typically cover the extreme forward angle region, where CLAS has a hole for the beam-dump. The only kinematics where a direct comparison is possible is between the CLAS g11a data and the CEA-Elings results at $\kcos \sim 0.81$, as shown in Fig.~\ref{fig:klam_ksig_clas_leps_cea}. We have also overlaid in this plot, the latest LEPS forward-angle data that are in excellent agreement with CLAS. Fig.~\ref{fig:clas_slac_scaled_klam_ksig} shows the comparison of $s^2 \times d\sigma/dt$ scaled cross sections between CLAS and SLAC-Boyarski -- while the shapes agree very well, the disagreement in scale is clearly visible.

\subsection{$\peta$}

\begin{figure}
  \includegraphics[width=3.3in]{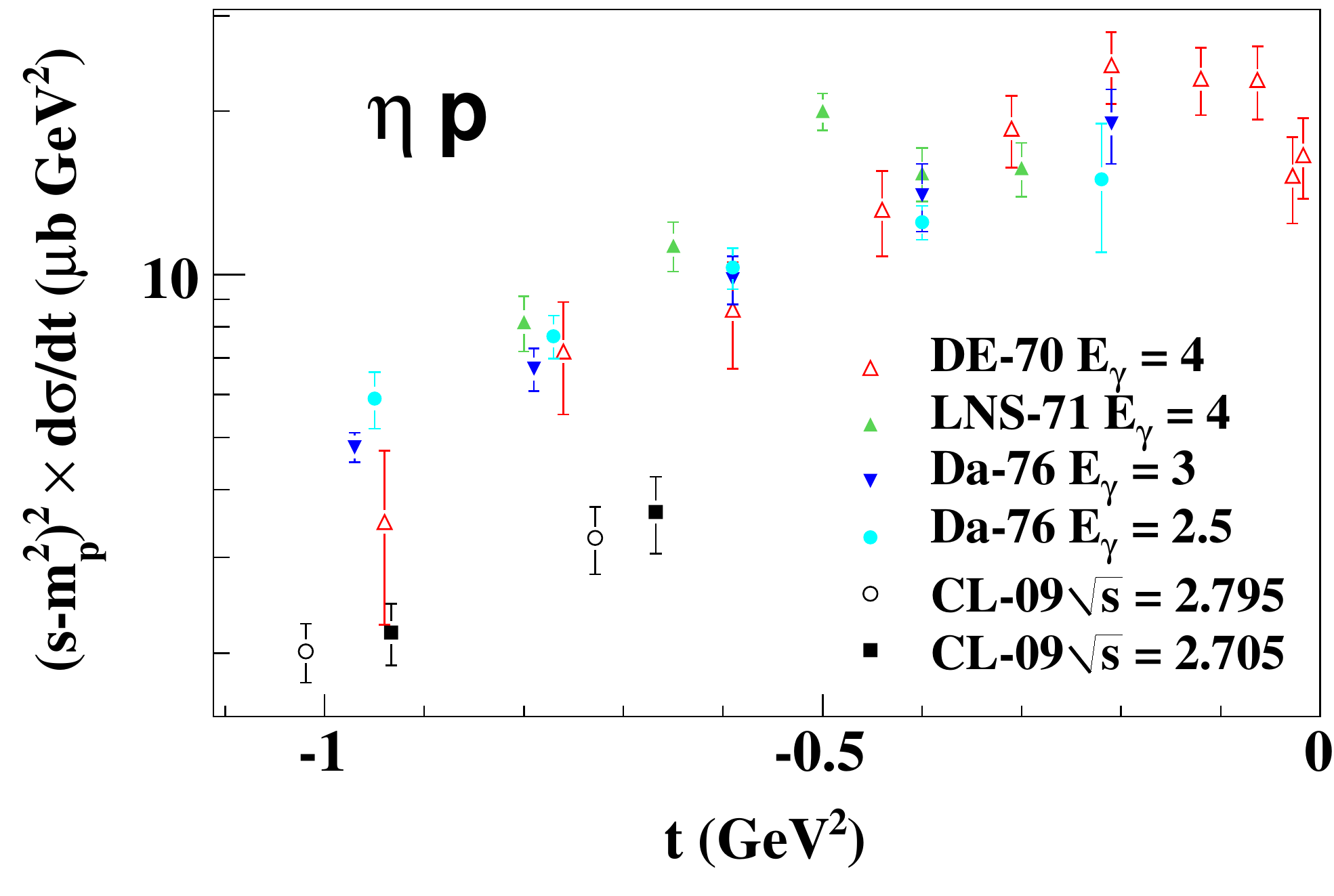}
  \caption[]{\label{fig:scaled_eta} (Color online) Comparison between the CLAS g11a~\cite{prc_eta} (CL-09), DESY-Braunschweig-1970~\cite{eta_desy} (DE-70), LNS-Dewire-1971~\cite{eta_lns_dewire} (LNS-71) and Daresbury-Bussey-1976~\cite{bussey_daresbury} (Da-76) scaled cross section results for $\peta$. The energies are listed in $E_\gamma$ for DESY/LNS/Daresbury and $\sqrt{s}$ for CLAS, in units of GeV. Both sets of results (DESY/LNS/Daresbury and CLAS) agree to a $d\sigma/dt \propto 1/(s-m_p^2)^2$ behavior as depicted by the tightly bunched set of points, but differ by a normalization factor. The error bars represent the statistical and systematic uncertainties added in quadrature.}
\end{figure}

\begin{figure}
  \includegraphics[width=3.3in]{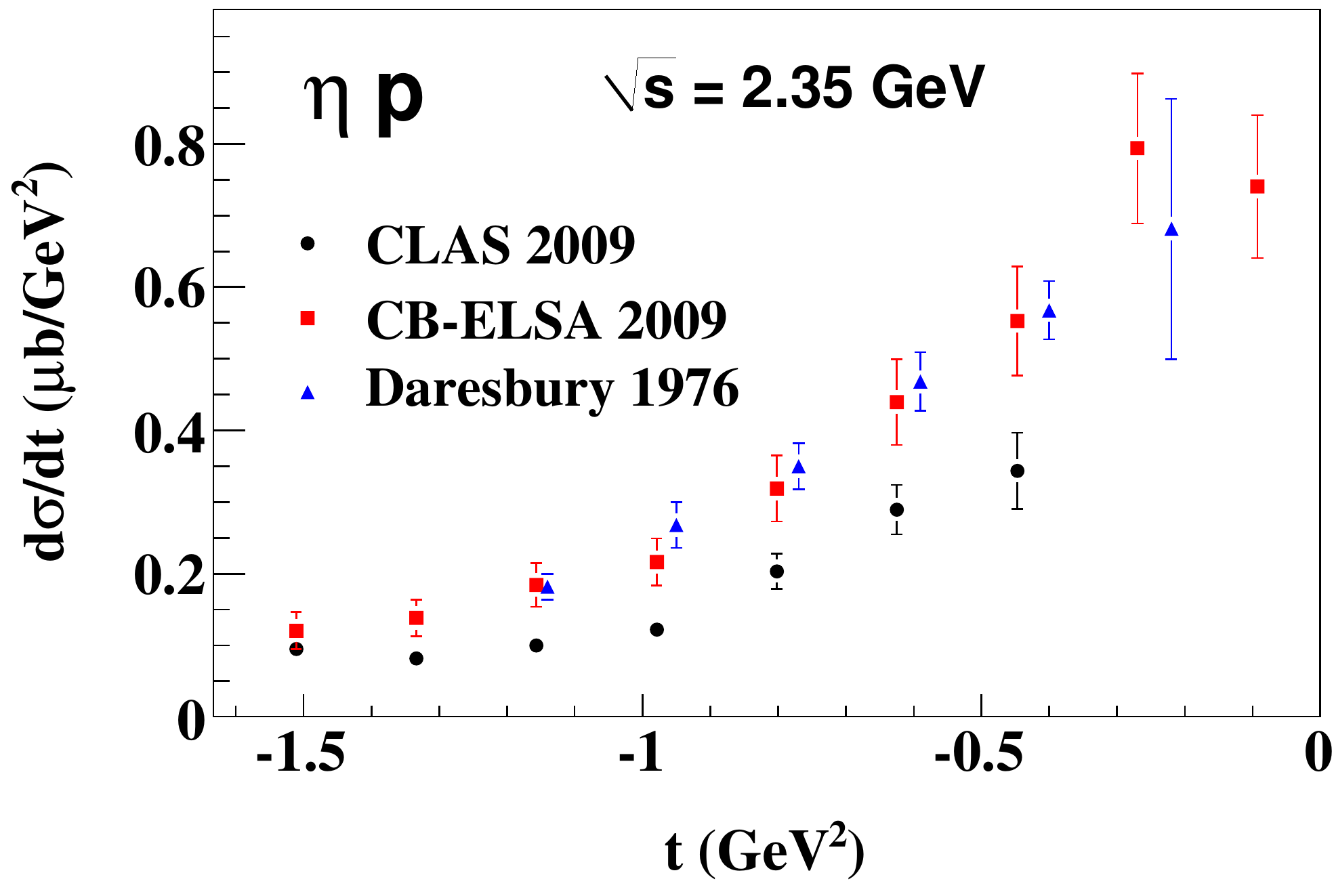}
  \caption[]{\label{fig:peta_clas_cbelsa_daresbury} (Color online) Comparison between the CLAS g11a~\cite{prc_eta}, CB-ELSA-2009~\cite{crede_eta_2009} and Daresbury-Bussey-1976~\cite{bussey_daresbury} for the $\peta$ channel in the energy bin $\sqrt{s} \approx 2.35$~GeV. CLAS appears to be systematically lower than both Daresbury and ELSA at forward-angles. The error bars represent the statistical and systematic uncertainties added in quadrature.}
\end{figure}

\begin{figure}
  \includegraphics[width=3.3in]{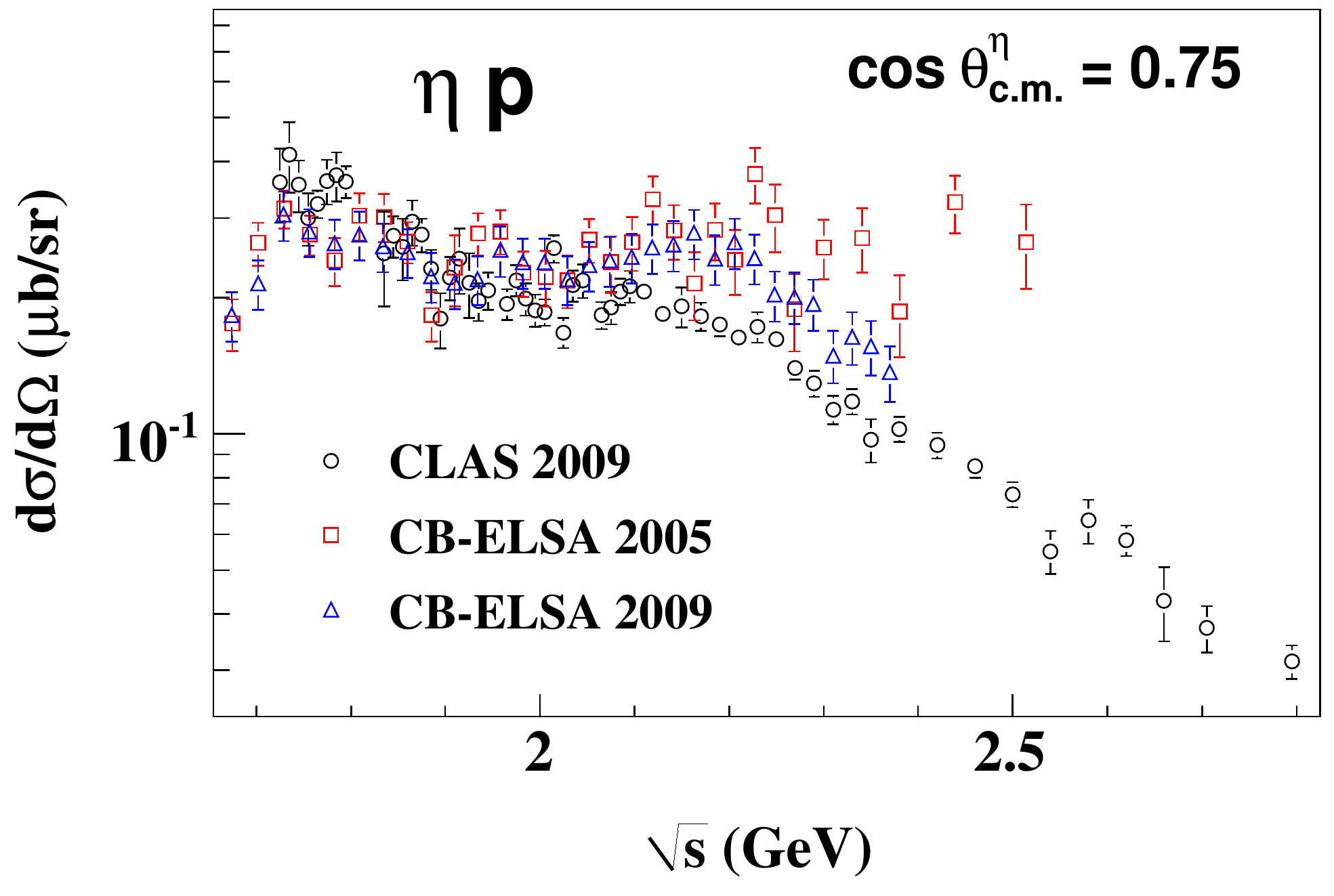}
  \caption[]{\label{fig:clas_elas_2005_elsa_2009} (Color online) Comparison between the CLAS g11a~\cite{prc_eta}, CB-ELSA-2009~\cite{crede_eta_2009} and CB-ELSA-2005~\cite{crede_eta_2005} for the $\peta$ channel in a forward-angle bin. Note that above $\sqrt{s}\approx 2.1$~GeV, $d\sigma/d\Omega$ vs. $\sqrt{s}$ appears flat in ELSA-2005, but in the ELSA-2009 version the cross section starts falling off with $\sqrt{s}$, as seen in the CLAS results. The error bars represent the statistical and systematic uncertainties added in quadrature.}
\end{figure}

The older high-energy forward-angle cross section data for this channel comprise of the following: CEA-Bellenger-1969~\cite{eta_cea_bellenger}, DESY-Braunschweig-1970~\cite{eta_desy}, LNS-Dewire-1971~\cite{eta_lns_dewire}, SLAC-Anderson-1971~\cite{anderson_slac} and Daresbury-Bussey-1976~\cite{bussey_daresbury}. As in the case of the $KY$ channels, a smooth $d\sigma/dt \propto 1/(s-m_p^2)^2$ behavior is reported in all these datasets that are also in fair to good agreement with each other. As shown in Fig.~\ref{fig:scaled_eta}, the CLAS data also shows a scaling behavior. However, while the scaled DESY/LNS/Daresbury cross-sections agree with each other, CLAS appears to be systematically lower. We also note here that none of the older experiments used a tagged beam.

In the previous sub-section, for the hyperons, we pointed out that CLAS is is good agreement with another recent experiment (LEPS) that had comparable statistical precision and employed a tagger photon beam. A perplexing issue for the $\peta$ channel is that the CLAS results do not agree with recent CB-ELSA measurements. Fig.~\ref{fig:peta_clas_cbelsa_daresbury} shows the comparison between CLAS~\cite{prc_eta}, CB-ELSA-2009~\cite{crede_eta_2009} and Daresbury-Bussey-1976~\cite{bussey_daresbury} in the energy bin $\sqrt{s} \approx 2.35$~GeV. At forward-angles, CB-ELSA and Daresbury agree well with each other, while CLAS is systematically lower. This issue was also mentioned in the work by Sibirtsev {\em et al.}~\cite{sibirtsev_eta_discrepancy}. While we do not have a resolution for the CB-ELSA/CLAS discrepancy at the moment, we make two comments on the issue. First, the CLAS g11a results were found to be in fair agreement with an earlier (unpublished) g1c analysis, pointing towards internal constistency within CLAS. Second, on the other hand, the recent ELSA-2009~\cite{crede_eta_2009} results show a marked difference from ELSA-2005~\cite{crede_eta_2005} in the foward-angle region above $\sqrt{s}\approx 2.1$~GeV. In the ELSA-2005 version, $d\sigma/d\Omega$ vs. $\sqrt{s}$ appears almost flat, while in the ELSA-2009 version, $d\sigma/d\Omega$ starts falling off with $\sqrt{s}$, as seen by CLAS. In the ELSA-2009 paper~\cite{crede_eta_2009}, this shift is attributed to an underestimated background at forward-angle and high $\sqrt{s}$ in ELSA-2005. While ELSA-2009 is still higher than CLAS, it is encouraging to see that at least the cross-section shape is in better agreement between the two datasets.

\subsection{$\piN$}

As mentioned in the introduction, since pions are most copiously produced in hadronic reactions, the single pion channels are where most of the world data reside. The $\piN$ channels have also been commonly used as the ``normalization channels'' between different world datasets. On many occassions, the $K^+ Y$ ($Y = \Lambda, \Sigma^0$) and $\peta$ results came from parasite analyses from an original $\pi^+ n$ and $\pi^0 p$ dataset. As an example, the high energy forward-angle SLAC-Boyarski $\pipN$~\cite{boyarski_pipN} results were first published in 1968, followed by the $K^+ Y$ results in 1969~\cite{boyarski_hyp}. The other relevant $\piN$ world data in this kinematic regime include CEA-Joseph-1967~\cite{joseph}, CEA-Elings-1967~\cite{elings}, DESY-Buschhorn-1966~\cite{buschhorn_66}, DESY-Buschhorn-1967~\cite{buschhorn_67}, DESY-Heide-1968~\cite{heide}, DESY-Braunschweig-1967~\cite{pi0n_braunschweig}, {\em et al}. Within themselves, agreement between these older results is fair to good.

\begin{figure}
  \includegraphics[width=3.3in]{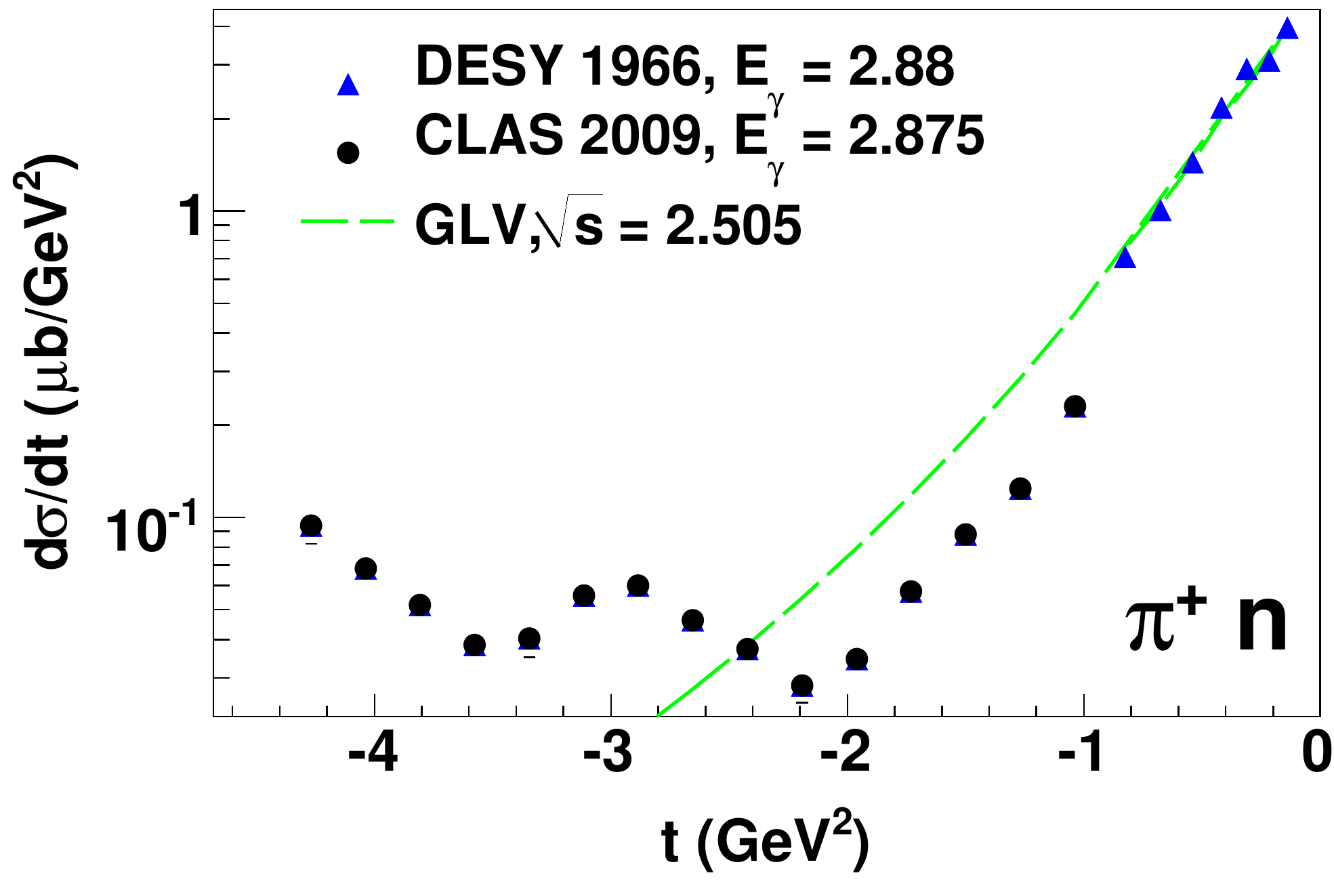}
  \caption[]{\label{fig:dugger_buschhorn} (Color online) Overlaid CLAS-Dugger-2009~\cite{dugger_pipN} and DESY-Buschhorn-1966~\cite{buschhorn_66} for the $\pi^+ n$ channel. The error bars represent the statistical and systematic uncertainties added in quadrature. The dashed green lines show a Regge-model prediction that fits well to the DESY data, but the projected model over-predicts the cross-sections at the CLAS kinemtics. We underscore the fact that this is not a direct comparison, although a scale discrepancy is ``plausible'' assessment.}
\end{figure}

\begin{figure}
  \includegraphics[width=3.3in]{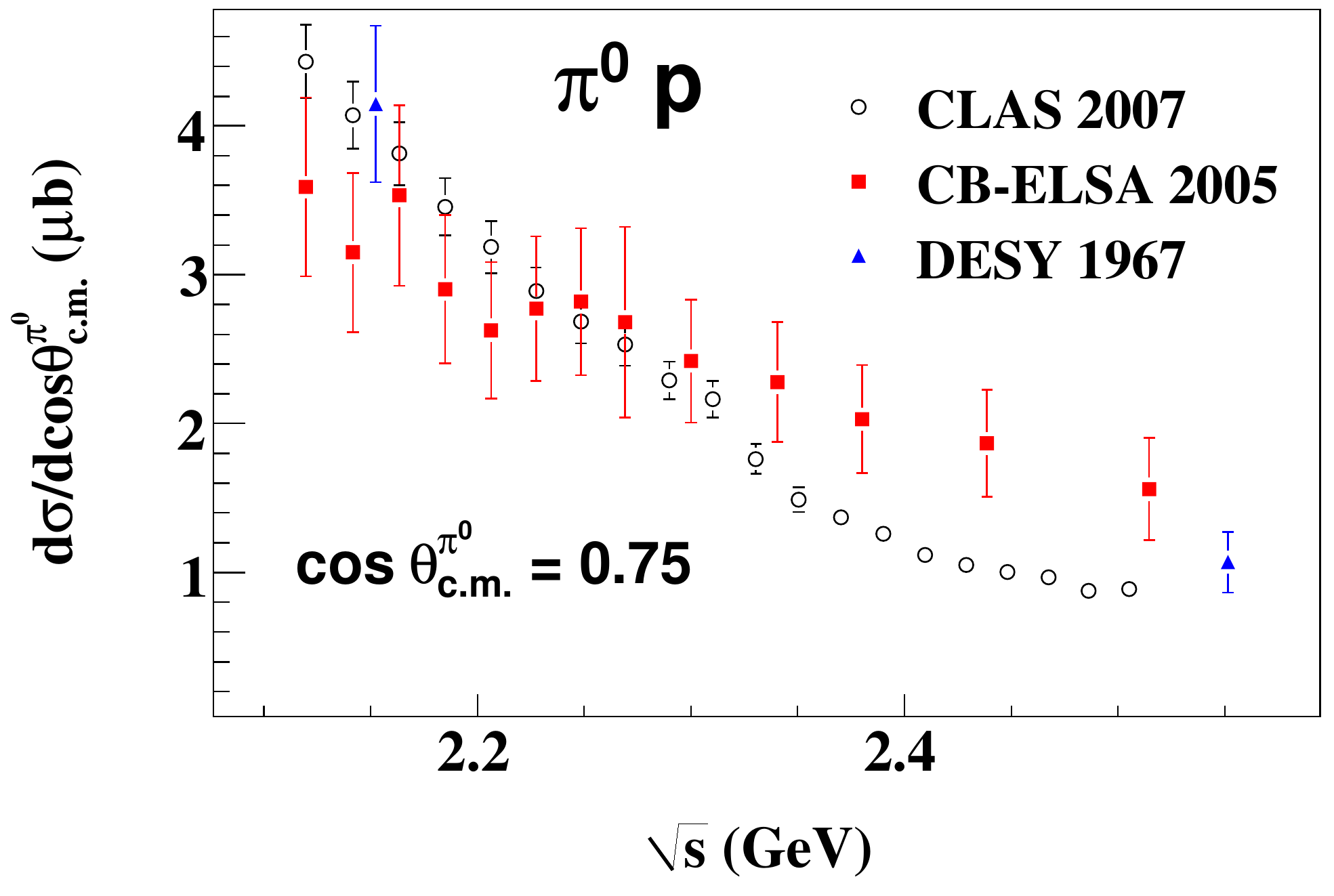}
  \caption[]{\label{fig:pi0p_clas-g1c_cbelsa_desy} (Color online) Comparison between the CLAS-Dugger-2009~\cite{dugger_pi0N}, CB-ELSA-Bartholomy-2005~\cite{pi0n_bartholomy} and DESY-Braunschweig-1967~\cite{pi0n_braunschweig} results for the $\pi^0 p$ channel. At higher energies and forward-angles, there is a scale discrepancy, most prominently visible between the CLAS and CB-ELSA results. The error bars represent the statistical and systematic uncertainties added in quadrature.}
\end{figure}

Since the trigger setting for the CLAS g11a experiment required detection of at least two charged particles, the single-pion channels were not accessible here. The CLAS g1c $\pi^+ n$~\cite{dugger_pipN} and $\pi^0 p$~\cite{dugger_pi0N} results go till about $\sqrt{s} \approx 2.55$~GeV, and have restricted coverage in the forward angles. Fig.~\ref{fig:dugger_buschhorn} shows the DESY-Buschhorn-1966~\cite{buschhorn_66} and CLAS-Dugger-2009~\cite{dugger_pipN} at $E_\gamma \approx 2.88$~GeV together. Unfortunately, the two datasets do not have a direct overlap. To guide the eye, we have overlaid a Regge-model prediction based on the work by Guidal-Laget-Vanderhaeghen (GLV)~\cite{glv}. The projected model prediction at the CLAS energies seems to hint that a scale discrepancy exists for the $\pi^+ n$ channel as well, though we underscore the fact that this is not a direct evidence. A direct comparison between CLAS and the older high-energy forward-angle SLAC/CEA/DESY data is also not possible in the $\pi^0 p$ sector due to non-overlapping kinematics. However, there are previous CB-ELSA~\cite{pi0n_bartholomy} data that overlap with the CLAS kinematics. Fig.~\ref{fig:pi0p_clas-g1c_cbelsa_desy} shows a comparison between CLAS-Dugger-2009~\cite{dugger_pi0N}, CB-ELSA-Bartholomy-2005~\cite{pi0n_bartholomy} and DESY-Braunschweig-1967~\cite{pi0n_braunschweig} for the $\pi^0 p$ channel. As for the $\peta$ channel, CB-ELSA results show a scale discrepancy with CLAS (and possibly a shape discrepancy as well).

\subsection{$\pomega$ and $\pphi$}

\begin{figure}
  \includegraphics[width=3.3in]{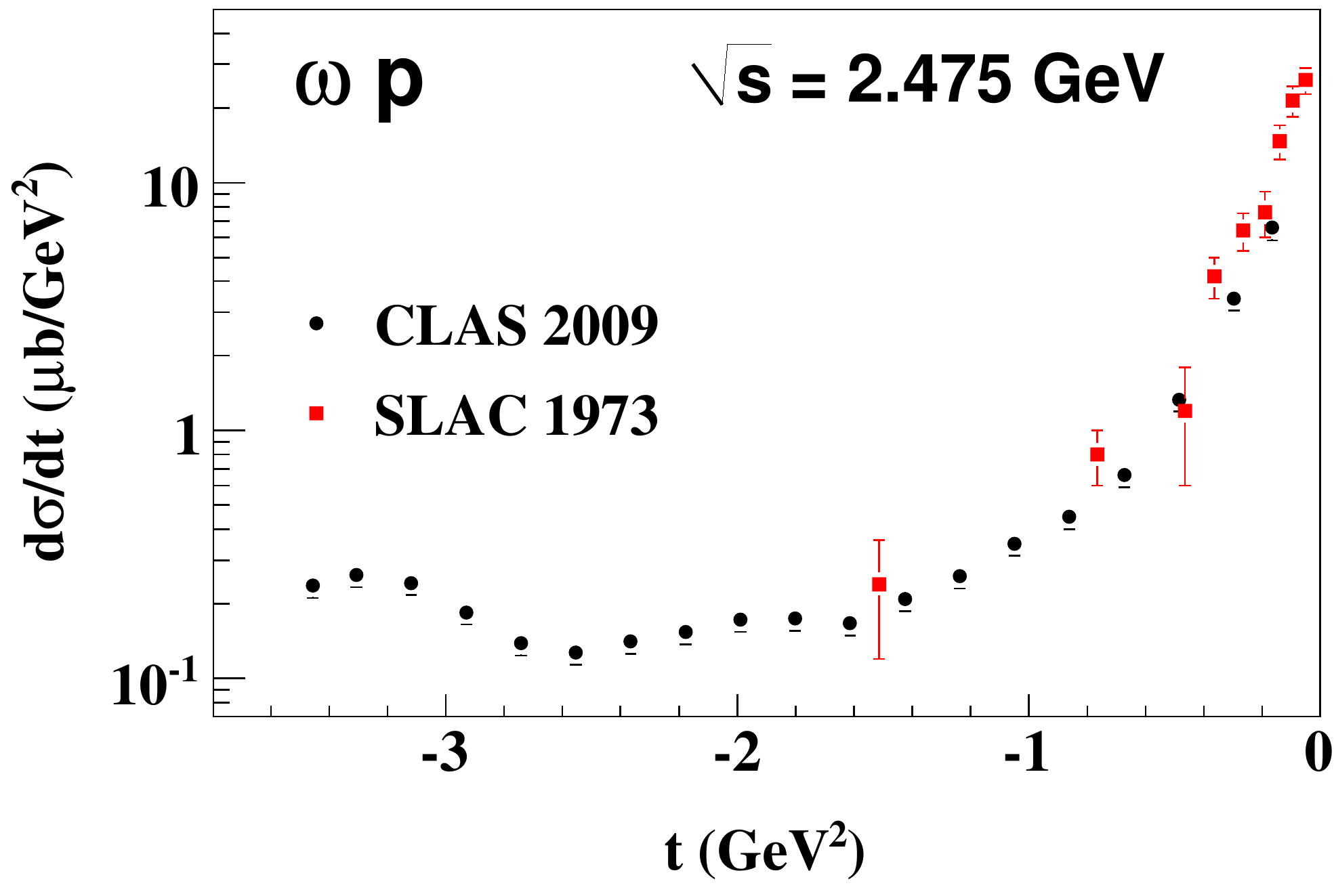}
  \caption[]{\label{fig:omega_clas_slac} (Color online) Comparison between the CLAS g11a~\cite{prc_omega} and SLAC-Ballam-1973~\cite{ballam} results at ${\sqrt{s} = 2.475}$~GeV for the $\pomega$ channel. The agremment is fair. The error bars represent the statistical and systematic uncertainties added in quadrature.}
\end{figure}

\begin{figure}
  \includegraphics[width=3.3in]{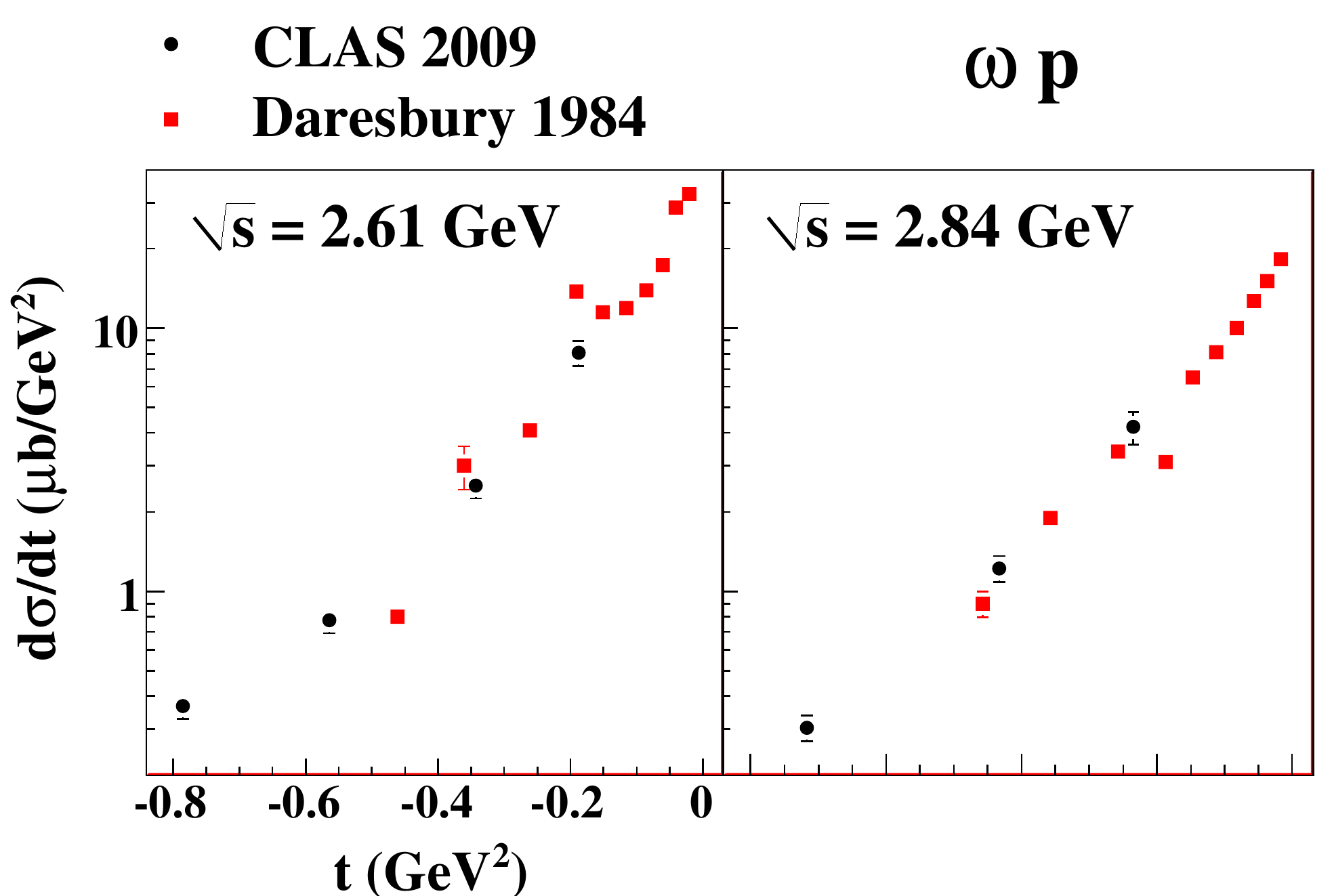}
  \caption[]{\label{fig:omega_clas_daresbury} (Color online) Comparison between the CLAS g11a~\cite{prc_omega} and Daresbury-Barber~\cite{barber_omega} results for the $\pomega$ channel. The agremment is fair. The error bars represent the statistical and systematic uncertainties added in quadrature.}
\end{figure}

\begin{figure}
  \centering
  \hspace{-3cm} \includegraphics[width=3.3in]{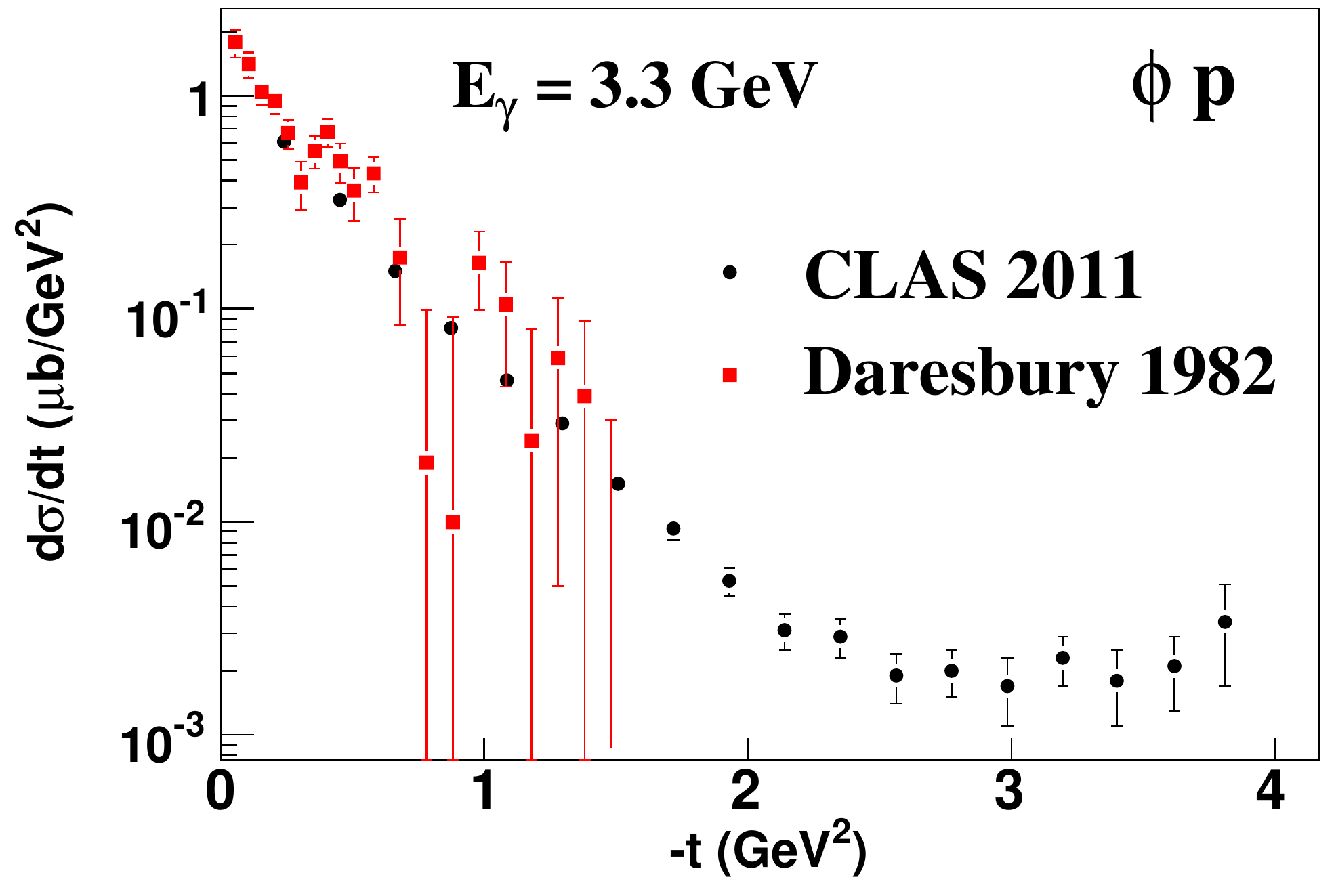} \hspace{-5cm} \vspace{-3cm}\rotatebox{45}{PRELIMINARY} \vspace{3cm}
  \caption[]{\label{fig:phi_clas_daresbury} (Color online) Comparison between the CLAS g11a~\cite{prc_phi} and Daresbury-Barber~\cite{barber_phi} results for the $\pphi$ channel. The Daresbury results have large error bars at higher $|t|$, but otherwise appear to be consistent with CLAS. The error bars represent the statistical and systematic uncertainties added in quadrature.}
\end{figure}

For the vector meson channels, $\pomega$ and $\pphi$, most the previous world data reside only in the diffractive region (large $s$ and $|t| \to 0$) and data at CLAS kinematics is scarce. The two previous vector meson world datasets that we compare the CLAS results to are SLAC-Ballam-1973~\cite{ballam} and Daresbury-Barber~\cite{barber_omega, barber_phi}. Figs.~\ref{fig:omega_clas_slac} and \ref{fig:omega_clas_daresbury} show the CLAS-SLAC cnd CLAS-Daresbury comparisons, respectively, for the $\pomega$ channel. Fig.~\ref{fig:phi_clas_daresbury} shows the comparison between CLAS and Daresbury for $\pphi$. The older data had wide energy bins while CLAS haw a much finer 10-MeV-wide $\sqrt{s}$ binning, the CLAS results are shown at the approximate bin-centers of the SLAC and Daresbury results. Within the shown error bars, the agreement is fair to good.

\section{Internal consistency checks within CLAS g11a}

As already mentioned, in the regions of overlapping kinematics, the results from g11a and g1c are in excellent agreement with each other. Given the very different analysis tools, trigger settings and acceptance calculation methods employed for the two experiments, this agreement is very encouraging. In the next step, several internal checks were run within g11a to ensure that the systematics were being understood well enough. The calculation of the differential cross section comprises of three elements -- yield extraction, detector acceptance and normalization (photon flux and target characterestics). If we assume that CLAS $\pomega$ results are correct, since they match very well with SLAC and Daresbury, we can rule out any discrepancy arising from the normalization part in the cross section calculation, since this is completely channel independent.

To check the reliability of our acceptance calculation, several steps were taken. The CLAS detector has a forward angle hole for the beam-dump. It is convievable that our understanding of the detector acceptance worsens as we approach the forward angle region. However, as shown in Fig.~\ref{fig:klam_ksig_clas_leps_cea} our agreement with LEPS, which is a dedicated forward-angle detector facility, shows that this is probably not the case. The magnetic setting for the CLAS spectrometer results in negatively charged tracks (corresponding to $K^-$ or $\pi^-$) being bent inwards toward the beam-dump hole. Therefore, the CLAS acceptance is generally lower for negatively tracks as compared to the positive ones. To check for possible shortcomings in our understanding of the $K^-$/$\pi^-$ acceptances, different reaction topologies were analyzed that included/excluded the detection of negative tracks. A summary of these topologies for each channel within CLAS g11a is listed in Table~\ref{table:topologies}. There are several noteworthy points here. First, any topology having more than one undetected final-state particle could not make use of the kinematic fitter. This pertains to the two-track topologies for the $\ksig$ and $\peta$ channels. The fact that the two- and three-track topologies agree with each other for these channels therefore indirectly strengthens our faith in the kinematic fitting procedure. Second, our partial wave analysis based physics-weighted acceptance calculation method required knowledge of all the final-state 4-momenta and was therefore limited to fully exclusive or a single final-state missing particle topologies. Since this was not available for the two-track $\ksig$ and $\peta$ topologies, a flat phase-space Monte Carlo event generator was used here. In fact, recall from Table~\ref{table:g11a_g1c}, a phase-space Monte Carlo generator was also used in the case of the g1c analyses. If the energy bin-width is small enough such that the physics does not vary too much across the bin and the break-up energies of the decays are small enough, a flat phase-space Monte Carlo should suffice. The fact that the different topologies agree well with each other shows that our understand of the CLAS accepatnce (in the fiducial regions) os reliable. Third, given that the $\pomega$ and $\peta$ three-track topologies involve the same set of detected final-state particles, it would be somewhat surprising if our acceptances for the $p$, $\pi^+$ and $\pi^-$ tracks are well understood for the $\pomega$ case, but not for $\peta$. A similar argument could be made between the $\pphi$ charged-mode and $\klam$ two-track topologies where the only difference is that the undetected particle is $K^-$ and $\pi^-$, respectively.

\begin{table}
  \centering
  \begin{tabular}{|c|c|c|}
  \hline
  \hline
  \multicolumn{3}{|c|}{CLAS g11a Reaction Topologies} \\
\hline
Channel & Topology & Name \\ \hline
\multirow{2}{*}{$\klam \to $} & $p K^+ \pi^-$ & three-track \\
                                      & $p K^+ (\pi^-)$ & two-track \\
\hline
\multirow{2}{*}{$\ksig \to $} & $\klam \gamma_f \to p K^+ \pi^-(\gamma_f)$ & three-track \\
                                      & $\klam \gamma_f \to p K^+ (\pi^- \gamma_f)$ & two-track \\
\hline
\multirow{2}{*}{$\peta \to $} & $p \pi^+ \pi^- (\pi^0)$ & three-track \\
                                      & $p \pi^+ (\pi^- \pi^0)$ & two-track \\
\hline
\multirow{1}{*}{$\petap \to $} & $p \pi^+ \pi^- (\eta)$ & three-track \\
\hline
\multirow{1}{*}{$\pomega \to $} & $p \pi^+ \pi^- (\pi^0)$ & three-track \\
\hline
\multirow{2}{*}{$\pphi \to $} & $p K^+ (K^-)$ & charged-mode \\
                              & $p K^0_S (K^0_L) \to p \pi^+ \pi^- (K^0_L)$ & neutral-mode \\
\hline
\end{tabular}
 \caption[]{\label{table:topologies} The reaction topologies for the six channels from the CLAS g11a experiment. Since a 2-prong trigger was used here, at least two charged particles were required to be detected. The undetected final-state particle(s) is(are) shown within parentheses for each case.}
\end{table}

\section{Tagged and untagged photon experiments}

One of the recurring aspects of the older SLAC/DESY/CEA experiments (for the pseudoscalar channels) is that they did not have a tagged photon beam. The energy bin-widths in these results were large ($\Delta E_\gamma \sim \mathcal{O}(1~\mbox{GeV})$) and with large uncertainties in the quoted $E_\gamma$. Aside from the fact that with a wide binning, the cross-section can vary considerably across the bin-wdith, another concern is that the conversion between the variables $t$ and $\cos \theta_{\mbox{\scriptsize c.m.}}^{\mbox{\scriptsize meson}}$, which depends on $E_\gamma$. Some of the older experiments even quote the electron beam energy as the estimated photon beam energy. It is also possible that the older experiments had been normalized to each other, which could explain their mutual agreement. In fact, the CLAS g11a data is the first photoproduction dataset that ``bridges'' the higher energy forward-angle regime (where discrepancies exist) with the lower energy regime (where there is no apparent discrepancy). The only problem with this explanation is that the CB-ELSA $\pi^0 p$~\cite{pi0n_bartholomy} and $\eta p$~\cite{crede_eta_2009, crede_eta_2005} data, which is more recent, and did use a tagged photon beam, also differs from CLAS at the forward-angles. Therefore, the CLAS-ELSA discrepancy remains an outstanding issue that needs to be resolved by future experiments.

For the two vector meson (channels $\pomega$ and $\pphi$) where we do not find a discrepancy, the SLAC-Ballam~\cite{ballam} data did not use a bremsstrahlung beam. In this case, a nearly mono-chromatic linearly polarized photon beam was produced by a collimated laser-backscattering procedure. The Daresbury-Barber~\cite{barber_omega, barber_phi} used a tagged bremsstrahlung beam.

\section{Effect on the hadrodynamic coupling constants}

Before proceeding further, we first clarify that our aim in this section is not to investigate photoproduction mechanisms {\em per se}  (a full coupled-channel partial wave analysis will be published separately). Rather, we would like to show that irrespective of the physics model chosen, a normalization discrepancy at the higher energies will have a significant bearing on future PWAs in search for resonances at lower energies.

We follow the Regge-based work outlined in GLV~\cite{glv} and adopted by the Ghent-RPR group~\cite{rpr_lambda,rpr_sigma} and others~\cite{rpr_eta}, and limit our discussion to the three channels $\pi^+ n$, $K^+ \Lambda$ and $K^+ \Sigma^0$. The first indication of a normalization problem between CLAS and SLAC can be found in the CLAS g1c $\klam/\ksig$ paper~\cite{bradford-dcs} where the GLV model from a fit to the SLAC high-energy results and projected down to CLAS energies consistently overpredicts the $\klam$ cross-sections (see Fig.~20 in Ref.~\cite{bradford-dcs} for example). In the basic GLV model, there are just two Reggeized $t$-channel exchange processes, a pseudo-scalar ($J^P = 0^-$) exchange and a vector meson ($J^P = 1^-$) exchange. GLV also assumes these exchange processes to be strongly degenerate with the higher spin exchanges (lying on the same Regge trajectory). As a result, incorporating just the lowest spin exchanges can suffice, and the model does a reasonable job in predicting the high energy data. The issue of degeneracy has been re-visited by Yu {\em et al.} in their recent work~\cite{yu_pion, yu_kaon}, where, instead of a strong degeneracy (both the couplings and the phases are degenerate), they assume a weak degeneracy (only the phases are degerate). In the Yu work, higher spin tensor exchanges are taken into account as a simple extension of the basic GLV model.

As such, we do not have any complaints about the physics motivation in this extended-GLV model of Yu. The problem is that Yu claims that within the Regge framework, addition of the tensor exchanges can lead to a reduction of the projected $\klam$ cross-sections at the CLAS energies, and therefore, these tensor exchanges are {\em necessary}. We also note that the Yu article does not incorporate the CLAS g11a data, which has a much closer kinematic proximity to the SLAC data. Once the CLAS g11a results are taken into account, it becomes amply clear that the problem lies not in the model, but in the data itself, as we showed in Sec.~\ref{sec:klam_ksig_discrepancy}. Therefore, attempts to reconcile both the CLAS and SLAC data within a single fit are essentially misleading. 

\subsection{Coupling constants}

To our knowledge, the only coupling that is relatively well known is $g_{\pi NN}$. Most authors place its value around 13 and GLV takes $g_{\pi NN} \approx 13.45$. Assuming a $20\%$ broken SU(3), GLV places the following limits on $g_{Kp\Lambda}$ and $g_{Kp\Sigma}$:
\begin{subequations}
\label{eqn:born_couplings}
\begin{eqnarray}
-16 \leq & g_{Kp\Lambda} &\leq -10.6 \\
3.2 \leq& g_{Kp\Sigma} &\leq 4.7.
\end{eqnarray}
\end{subequations}
These couplings enter via the Born terms, $t$-channel pion/kaon exchange and $s$-channel nucleon/hyperon exchange. The latter is required for restoration of gauge symmetry, since the sole $t$-channel Born term breaks gauge invariance. Aside from these Born terms, $t$-channel vector meson exchanges can also occur. The couplings for this case are much less well known. For example, $g_{\rho N N}$ varies from 1.9~\cite{eta_he_saghai} to 3.4, as taken by GLV. Similarly, $\kappa_{\rho NN}$ is found to vary between 1.5 and 6.6 in the literature; GLV takes $\kappa_{\rho NN} = 6.1$. Given $g_{\rho N N}$ and $\kappa_{\rho NN}$, it is possible to estimate $g_{K^\ast pY}$ and $\kappa_{K^\ast pY}$ for the hyperons ($Y$) using SU(3)~\cite{glv}:
\begin{subequations}
\label{eqn:vector_couplings}
\begin{eqnarray}
g_{K^\ast p\Lambda} = -6.08 && \kappa_{K^\ast p\Lambda} = +3.66 \\
g_{K^\ast p\Sigma} = -3.51 && \kappa_{K^\ast p\Lambda} = -1.22.
\end{eqnarray}
\end{subequations}
However, given how much uncertainty there is in the values of the $\rho$ couplings themselves, even assuming unbroken SU(3), it is clear that the $K^\ast$ couplings remain poorly known. Finally, we also point out that the values of these couplings are model-dependent. The isobar-models (instead of the Regge-based models) include phenomenological form-factors that bring in further model-dependence.

\subsection{The SLAC forward-angle shapes}

\begin{figure}
  \centering
  \includegraphics[width=3.3in]{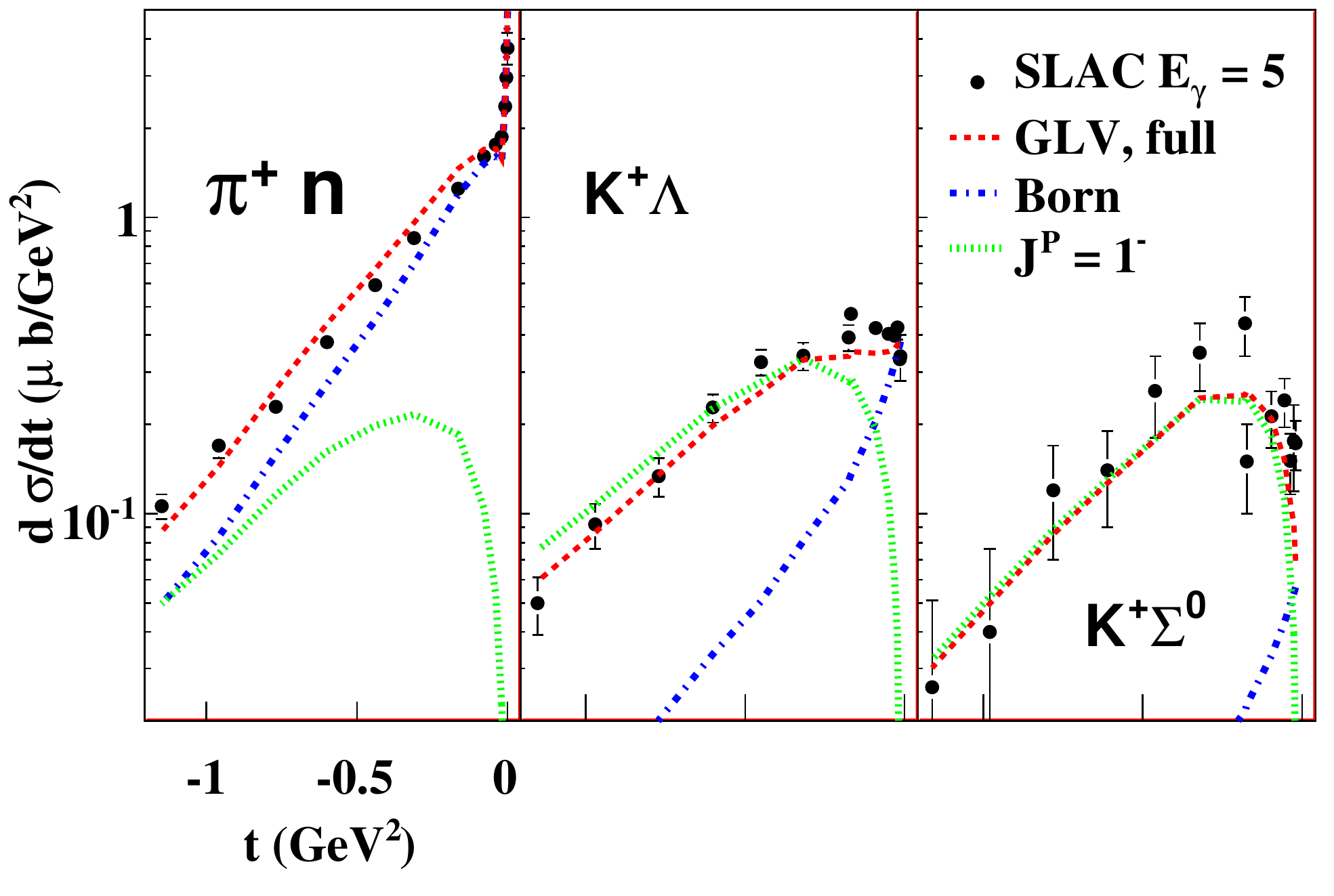}
  \caption[]{\label{fig:glv_slac_boyarski_shape} (Color online) Comparison between the forward-angle $d\sigma/dt$ for the $\pi^+ n$~\cite{boyarski_pipN}, $\klam$~\cite{boyarski_hyp} and $\ksig$~\cite{boyarski_hyp} channels from SLAC-Boyarski at $E_\gamma = 5$~GeV. At $|t| \to 0$, $\pi^+ n$ shows a rise, $\klam$ shows more of a plateau, while $\ksig$ shows a fall-off. The curves show the GLV~\cite{glv} Regge-model results. See text for details.}
\end{figure}

Fig.~\ref{fig:glv_slac_boyarski_shape} shows the forward-angle SLAC-Boyarski~\cite{boyarski_pipN,boyarski_hyp} results at $E_\gamma = 5$~GeV for the $\pipN$, $\klam$ and $\ksig$ channels overlaid with the GLV~\cite{glv} model fit results. The notable feature of interest we want to point out here is the $|t| \to 0$ behavior. For $\pipN$, the cross-section shows a steep peak; for $\klam$, this is more a flat plateau, while, for $\ksig$, it shows a fall off. In the GLV picture, this forward-angle behavior is dictated by the ratio between the Born term and the vector meson exchang term. The Born term consists of two pieces, the conventional $t$-channel $\pi^+$/$K^+$ exchange, plus, the electric part of the $s$-channel nucleon/hyperon exchange that is added to restore gauge invariance. The sharp rise for the $\pipN$ case is due to a large $g_{\pi NN}$ in the $s$-channel Born term. Relative to $g_{\pi NN}$, $g_{\rho NN}$ (for the vector-meson $J^P = 1^-$ exchange) is small, and therefore, the Born term dominates. In the case of the hyperon ($KY$) channels, since there is no sharp rise at $|t| \to 0$, this means, $g_{KpY}$ has to be relatively smaller than $g_{K^\ast p Y}$ and the ratio of these two couplings determines whether the $|t| \to 0$ should be a plateau ($\klam$) or a drop-off ($\ksig$). {\em At $|t| \approx 0$, the cross-section is completely fixed by $g_{KpY}$ in the Born term, since the $K^\ast$ exchange contribution vanishes here}.

Furthermore, it can be see from Fig.~\ref{fig:glv_slac_boyarski_shape} that in the case of the hyperons, away from $|t| = 0$, the natural-parity $K^\ast$ exchange term dominates, which would signify that the beam-asymmetry observable ($\Sigma$) be close to +1. This is indeed what is seen in the 16-GeV SLAC results~\cite{quinn}. 

Since the forward-angle coverage in the CLAS g11a results~\cite{prc_klam, prc_ksig} extends till $\kcos \leq 0.95$ only, these results do not contain information on the shape of the cross-section at $|t| \to 0$. Therefore, if we neglect the older SLAC results, ab initio, we do know whether the forward-angle cross-section is a rise, or plateau or a fall-off. On the other hand, the LEPS detector is a dedicated forward-angle detector, and as shown in Fig.~\ref{fig:klam_ksig_clas_leps_cea}, in regions where the kinematics overlap, the CLAS results are in excellent agreement with LEPS. The LEPS hyperon data~\cite{leps_klam_ksig_sumihama} in the forward-most angular bin show a slight fall-off in the cross-sections at $|t| \to 0$, in agreement with the SLAC-Boyarski results. Therefore, even if we do not include the SLAC data directly, it is plausible to incorporate the shapes of the forward-angle SLAC results. More specifically, we impose the restriction that the extrapolation of any PWA fit results into the $|t| \to 0$ region should not show a rise for the $\klam$ and $\ksig$ channels.  

\subsection{New fits incorporating the CLAS data}

\begin{figure}
  \centering
  \subfigure[]{
    {\includegraphics[width=3.5in]{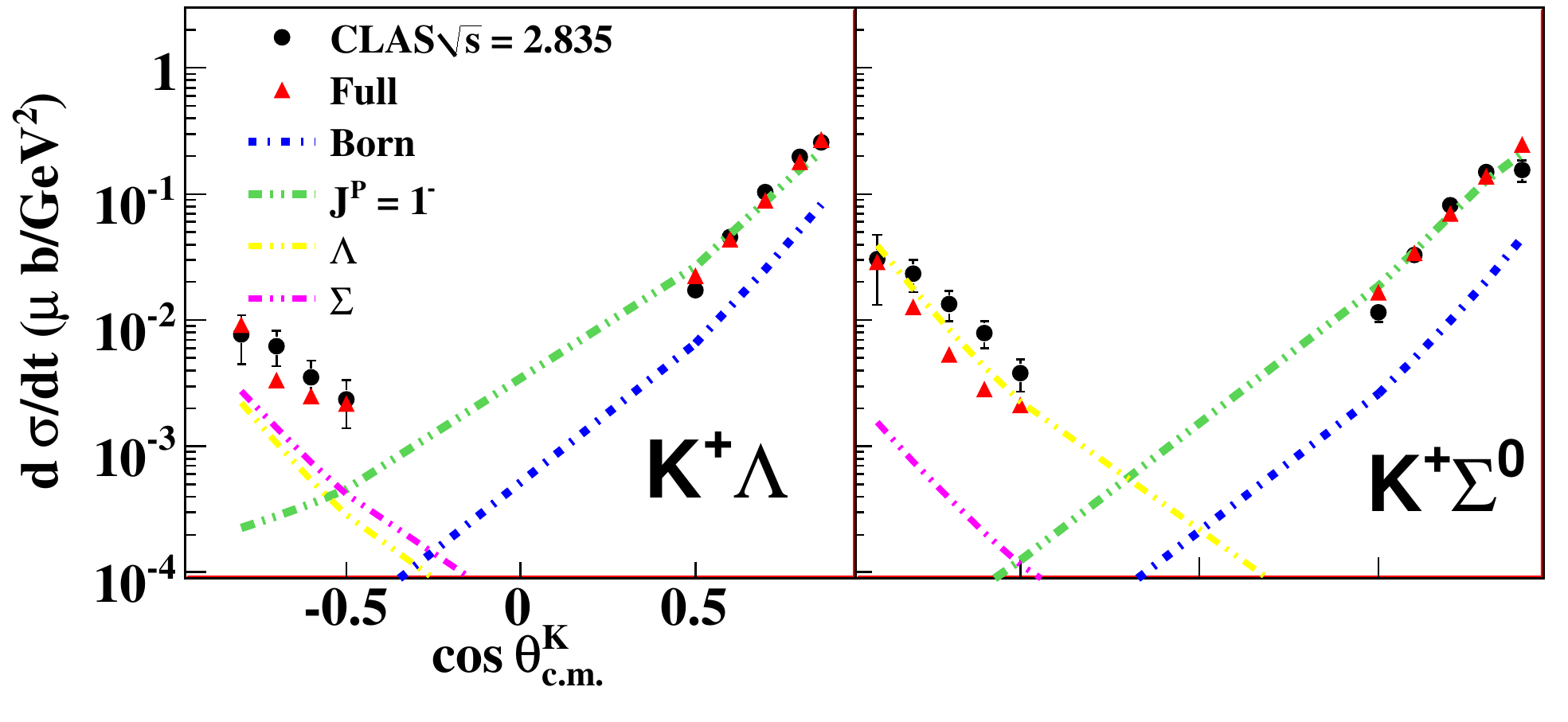}}
  }
  \subfigure[]{
    {\includegraphics[width=3.5in]{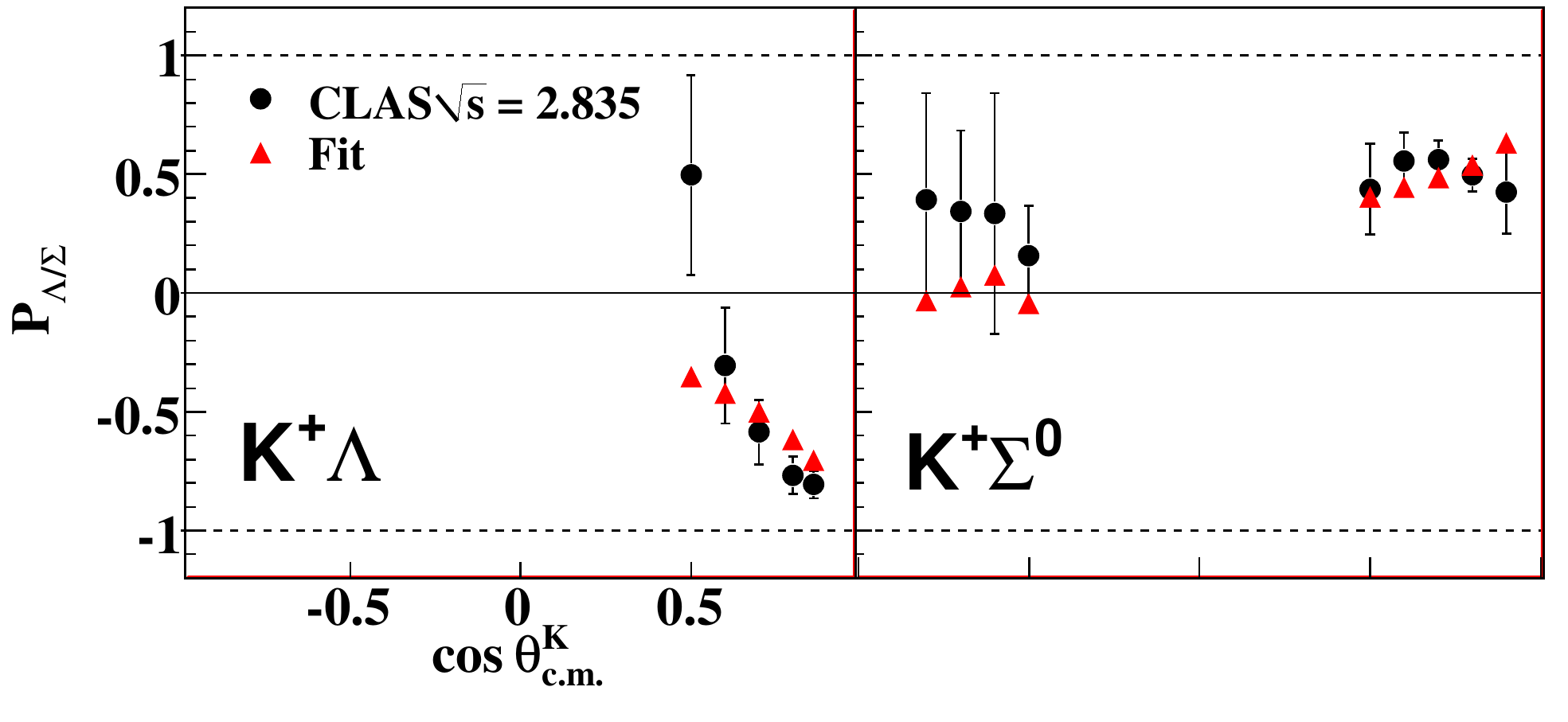}}
  }
  \subfigure[]{
    {\includegraphics[width=3.5in]{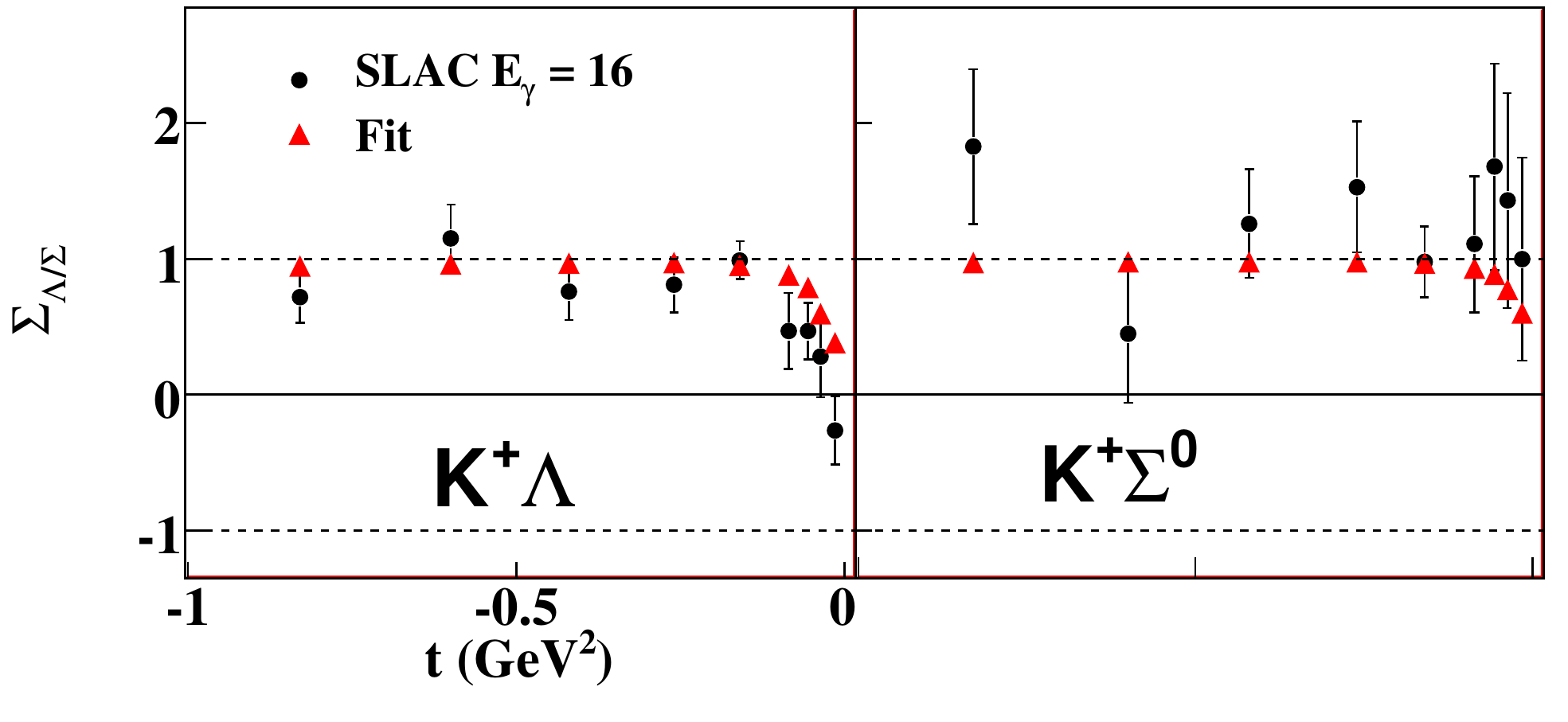}}
  }
\caption[]{\label{fig:new_fits} Results from a fit using the GLV model and incorporating all CLAS data points with $\sqrt{s} \geq 2.6$~GeV and $|\kcos| > 0.5$, as well as the $E_\gamma = 16$~GeV SLAC beam-asymmetry results~\cite{quinn}. }
\end{figure}

The results of our fit using the GLV model and all CLAS data points $\sqrt{s} \geq 2.6$~GeV (high energy), $|\kcos| > 0.5$ (forward- and backward-angle) as well as the SLAC $E_\gamma = 16$~GeV beam-asymmetry results~\cite{quinn} are shown in Fig.~\ref{fig:new_fits}. We assume a rotating phase for all the Regge amplitudes and the the $\klam$ and $\ksig$ channels were coupled together. The latter implies that the $u$-channel terms (involving $g_{Kp\Lambda}$ and $g_{Kp\Sigma}$) did not involve any new coupling. This adds to the internal consistency of our fit results. The values of the couplings we obtained are: $g_{Kp\Lambda} = -9.5$, $g_{Kp\Sigma} =5.6$, $g_{K^\ast p \Lambda} = -14.5$, $\kappa_{K^\ast p \Lambda} = 1.7$, $g_{K^\ast p \Sigma} = -14.5$ and $\kappa_{K^\ast p \Sigma} = -1.3$. We note here that the exact values of the couplings could depend on the choice of the phase choices and a more exhaustive study of the phases is currently underway . However, the values of the three couplings $g_{Kp\Lambda}$, $g_{K^\ast p \Lambda}$ and $g_{K^\ast p \Sigma}$ will certainly be lower than what was found in the original GLV work. The value of the $g_{Kp\Lambda}$ turns out to be especially important, since this contributes to the Born term in $\klam$, that dominates at near-threshold. Therefore, from the perspective of searches for $s$-channel resonances, it is very important that $g_{Kp\Lambda}$ be well-known. For $\ksig$, the Born-term plays a less dominant role, so that the contribution from $g_{Kp\Sigma}$ is small. However, the $K^\ast$ couplings $g_{K^\ast p \Lambda}$ and $g_{K^\ast p \Sigma}$ again play important roles.

\section{Discussion and future work}

In this article we have conducted a global survey of normalization issues in photoproduction reactions including several different reaction channels. Our chief goal is to highlight the fact that differential cross-sections from older SLAC/DESY/CEA data for the pseudoscalar meson channels that used untagged bremsstrahlung photon beams are roughly a factor of two larger than the more recent tagged photon experiments from CLAS and elsewhere. The older high energy SLAC data have been conventionally been used to ``fix'' the couplings for the non-resonant background processes. In light of the new CLAS results, these couplings will need to be re-visited and could potentially play an important in searches for $s$-channel resonances at lower energies. This is especially a matter of concern for the $\klam$ channel where the Born terms dominates at threshold.

Along with the CLAS/SLAC discrepancy, we also see a discrepancy between CLAS and ELSA data for the $\eta$ and $\pi^0$ channels at forward-angles. As of this writing, the CLAS/ELSA discrepancy is under further investigation. It is possible that newer precision data at higher energy and forward-angles for these channels would be required from other facilities like LEPS and GRAAL to settle this discrepancy issue. We hope these discrepancies will be resolved soon, since any scale ambiguities in the couplings will directly affect analysis of the ``complete'' experiments that forms a major goal for several facilities across the globe.

\end{document}